\documentstyle{mn}

\newif\ifAMStwofonts

\ifoldfss
  \ifCUPmtlplainloaded \else
    \NewTextAlphabet{textbfit} {cmbxti10} {}
    \NewTextAlphabet{textbfss} {cmssbx10} {}
    \NewMathAlphabet{mathbfit} {cmbxti10} {} 
    \NewMathAlphabet{mathbfss} {cmssbx10} {} 
  \fi
  \ifAMStwofonts
    \ifCUPmtlplainloaded \else
      \NewSymbolFont{upmath} {eurm10}
      \NewSymbolFont{AMSa} {msam10}
      \NewMathSymbol{\upi}     {0}{upmath}{19}
      \NewMathSymbol{\umu}     {0}{upmath}{16}
      \NewMathSymbol{\upartial}{0}{upmath}{40}
      \NewMathSymbol{\leqslant}{3}{AMSa}{36}
      \NewMathSymbol{\geqslant}{3}{AMSa}{3E}

      \let\geq=\geqslant 
    \fi
  \fi
\fi 

\ifnfssone
  \newmathalphabet{\mathit}
  \addtoversion{normal}{\mathit}{cmr}{m}{it}
  \addtoversion{bold}{\mathit}{cmr}{bx}{it}
  \newmathalphabet{\mathbfit} 
  \addtoversion{normal}{\mathbfit}{cmr}{bx}{it}
  \addtoversion{bold}{\mathbfit}{cmr}{bx}{it}
  \newmathalphabet{\mathbfss} 
  \addtoversion{normal}{\mathbfss}{cmss}{bx}{n}
  \addtoversion{bold}{\mathbfss}{cmss}{bx}{n}
  \ifAMStwofonts
    \ifCUPmtlplainloaded \else
      \UseAMStwoboldmath
      \makeatletter
      \new@mathgroup\upmath@group
      \define@mathgroup\mv@normal\upmath@group{eur}{m}{n}
      \define@mathgroup\mv@bold\upmath@group{eur}{b}{n}
      \edef\UPM{\hexnumber\upmath@group}
      \new@mathgroup\amsa@group
      \define@mathgroup\mv@normal\amsa@group{msa}{m}{n}
      \define@mathgroup\mv@bold\amsa@group{msa}{m}{n}
      \edef\AMSa{\hexnumber\amsa@group}
      \makeatother
      \mathchardef\upi="0\UPM19
      \mathchardef\umu="0\UPM16
      \mathchardef\upartial="0\UPM40
      \mathchardef\leqslant="3\AMSa36
      \mathchardef\geqslant="3\AMSa3E

      \let\geq=\geqslant 
    \fi
  \fi
\fi 

\ifnfsstwo
  \DeclareMathAlphabet{\mathbfit}{OT1}{cmr}{bx}{it}
  \SetMathAlphabet\mathbfit{bold}{OT1}{cmr}{bx}{it}
  \DeclareMathAlphabet{\mathbfss}{OT1}{cmss}{bx}{n}
  \SetMathAlphabet\mathbfss{bold}{OT1}{cmss}{bx}{n}
  \ifAMStwofonts
    \ifCUPmtlplainloaded \else
      \DeclareSymbolFont{UPM}{U}{eur}{m}{n}
      \SetSymbolFont{UPM}{bold}{U}{eur}{b}{n}
      \DeclareSymbolFont{AMSa}{U}{msa}{m}{n}
      \DeclareMathSymbol{\upi}{0}{UPM}{"19}
      \DeclareMathSymbol{\umu}{0}{UPM}{"16}
      \DeclareMathSymbol{\upartial}{0}{UPM}{"40}
      \DeclareMathSymbol{\leqslant}{3}{AMSa}{"36}
      \DeclareMathSymbol{\geqslant}{3}{AMSa}{"3E}

      \let\geq=\geqslant 
    \fi
  \fi
\fi 

\ifCUPmtlplainloaded \else
  \ifAMStwofonts \else 
    \def\upi{\pi}
    \def\umu{\mu}
    \def\upartial{\partial}
  \fi
\fi
\title{Smoothed Particle Hydrodynamics with particle splitting, 
applied to self-gravitating collapse}
\author[S. Kitsionas \& A. P. Whitworth]
       {S. Kitsionas,\thanks{email: spyrmyr@hotmail.com -- present address: 
        14 John Kennedy str., 16121 Kesariani, Athens, Greece}
        A. P. Whitworth\thanks{email: ant@astro.cf.ac.uk}\\
        Department of Physics \& Astronomy, Cardiff University, 
        PO Box 913, 5 The Parade, Cardiff CF24 3YB, Wales, UK}

\date{Accepted.
      Received;
      in original form}

\pagerange{\pageref{firstpage}--\pageref{lastpage}}
\pubyear{2001}

\newcommand{\cld}{_{\mbox{\small child}}}
\newcommand{\cri}{_{\mbox{\small crit}}}
\newcommand{\FF}{_{\mbox{\tiny FF}}}
\newcommand{\Jea}{_{\mbox{\small Jeans}}}
\newcommand{\nei}{_{\mbox{\small neib}}}
\newcommand{\pnt}{_{\mbox{\small parent}}}
\newcommand{\pea}{_{\mbox{\small peak}}}
\newcommand{\spl}{_{\mbox{\small split}}}
\newcommand{\ttl}{_{\mbox{\small total}}}
\newcommand{\hlf}{\frac{1}{2}}
\newcommand{\mxm}{_{\mbox{\small max}}}

\begin{document}

\maketitle

\label{firstpage}

\begin{abstract}

We describe and demonstrate a method for increasing the resolution 
locally in a Smoothed Particle Hydrodynamic (SPH) simulation, 
by splitting particles. We show that in simulations of self-gravitating 
collapse (of the sort which are presumed to occur in star formation) 
the method is stable, and 
affords great savings in computer time and memory. When applied 
to the standard Boss \& Bodenheimer test -- which 
has been shown to depend critically on fulfilment of the Jeans 
Condition -- the results are comparable both with those obtained 
using Adaptive Mesh Refinement, and with those obtained using a standard 
high-resolution SPH simulation, 
but they are achieved with considerably less computational resource. 
Further development and testing is required before the method can safely 
be applied to more general flows.

\end{abstract}
   

\section{Introduction}

Numerical simulations of star formation {\bf --} even the highly 
idealized simulations to which contemporary astrophysics is 
limited {\bf --} require very large computational resources. 
The dynamics is self-gravitating, so the gravitational field has 
to be recalculated at each time-step. The ranges of density ($\sim 
10^{-22}$ g cm$^{-3}$ to $\sim 10^1$ g cm$^{-3}$) and linear 
scale ($\sim 10^{20}$ cm to $\sim 10^{10}$ cm) are very large, 
so high resolution is needed. The geometry is complex, so 
three-dimensional simulations are essential. The dynamics is 
affected by a variety of radiative, thermal, chemical and 
magnetic effects, with the result that the energy equation is 
not a local function of state. The initial and boundary 
conditions are chaotic and poorly constrained by observation.

Smoothed Particle Hydrodynamics (SPH) is a particle-based 
hydrodynamic scheme which accommodates some of these 
requirements automatically, by virtue of being Lagrangian, 
having no imposed geometrical constraints, and being readily 
combined with particle-based gravity solvers, for example 
Tree Code Gravity (TCG; Barnes \& Hut 1986; Hernquist 1987; Hernquist 
\& Katz 1989). SPH is also relatively straightforward to implement, at 
least in its most basic formulation. More refined versions have to 
be used to obtain realistic results when there are large gradients 
of density (e.g. shocks and ionization fronts) and/or velocity (e.g. 
shocks and rapidly shearing discs).

Since its initial realization (Lucy 1977; Gingold \& Monaghan 
1977), SPH has been developed and refined by several workers, 
for example Lattanzio et al. (1985; artificial viscosity), 
Nelson \& Papaloizou (1993, 1994; variable $h$), Bate, Bonnell \& 
Price (1995; sink particles), Watkins et al. (1995; Navier-Stokes 
viscosity), Morris \& Monaghan (1997; time-varying artificial 
viscosity), Nelson \& Langer (1997; scheme for treating the thermal 
physics), Owen et al. (1998; tensorial smoothing kernels), Inutsuka 
\& Imaeda (2001; Godunov scheme for interparticle hydro forces), Kessel-Deynet 
\& Burkert (2000; scheme for following ionization fronts).

The principle alternatives to SPH are finite difference (FD) codes, 
or hybrids which combine particles and cells. FD codes are much better 
at resolving large density and/or velocity gradients. They should 
probably be the method of choice for simulating flows with high Mach 
Number. However, in three dimensions they are not as easy to implement 
as SPH, and they are not Lagrangian. Moreover, it should be born in 
mind that FD codes benefit from having had very many more person-years 
of development than SPH.

In FD codes, the high local 
resolution required for calculations of protostellar collapse 
can be obtained by means of nested grids. These nested grids can 
either be introduced at the outset to cover places where it is 
anticipated that extra resolution will be required (e.g. 
Burkert \& Bodenheimer 1993); or they can be introduced and 
removed during the course of the simulation, whenever and 
wherever they are required, as in Adaptive Mesh Refinement 
(AMR; Berger \& Colella 1989, Truelove et al. 1997, 1998). 
With AMR, the resolution can, in principle, 
be increased indefinitely in regions where this is necessary. 

In standard SPH, the linear resolution is $\sim 
(m/\rho)^{1/3}$, where $m$ is the mass of an individual particle 
and $\rho$ is the local density. In this paper we describe a 
rather straightforward algorithm whereby the resolution of an SPH 
code can also, in principle, be increased indefinitely, by 
splitting particles. We test the method on the standard Boss 
\& Bodenheimer (1979) test, using the Jeans condition to 
trigger splitting. Particle splitting and merging were used by 
Meglicki, Wickramasinghe \& Bicknell (1993) in their simulations 
of accretion discs, and particle merging had been used earlier by 
Monaghan \& Varnas (1988) in simulations of interstellar cloud 
complexes. However, in these earlier implementations splitting 
was used to maintain resolution in low-density regions, and merging 
was used to avoid following large numbers of particles in high-density 
regions (essentially the motivation for sink particles; Bate, Bonnell 
\& Price, 1995). Therefore, 
the philosophy was very different from that adopted here, where high 
resolution is advocated in high-density regions to avoid violating the 
Jeans condition (i.e. to ensure resolution of the local Jeans mass). 
Moreover, no tests were reported in these earlier papers.

Section 2 gives a brief review of standard SPH. Section 3 
sketches the implementation of Particle Splitting, at the 
microscopic level, and Section 4 describes how the best 
parameters for Particle Splitting are determined. Sections 5 
and 6 describe two algorithms for Particle Splitting at the 
macroscopic level. In Nested Splitting (Section 5) all the 
particles in a prescribed sub-domain are split from some 
predetermined time $t\spl$ onwards, so that in effect a 
high-resolution simulation is performed inside the sub-domain using 
initial and boundary conditions supplied by the coarser simulation 
in the overall computational domain. In On-The-Fly Splitting 
(Section 6), particles are split only when this is necessary 
according to some local criterion such as the Jeans Condition. 
On-The-Fly Splitting is the more computationally efficient 
procedure, requiring no manual intervention and no prior 
knowledge of where high resolution will be required; it corresponds 
closely to AMR. Section 7 introduces the Jeans Condition, and 
Section 8 demonstrates how effectively and economically SPH 
performs the standard Boss \& Bodenheimer (1979) rotating collapse 
test when Particle 
Splitting is included. Section 9 discusses the results and compares 
them with those obtained using AMR. Our conclusions are summarized 
in Section 10.

\section{Basic Smoothed Particle Hydrodynamics and Tree-Code Gravity}

In SPH the evolution of the gas is predicted by following the 
motions of an ensemble of $i\ttl$ particles which act as discrete 
sampling points (e.g. Monaghan 1992). Each particle has associated 
with it a mass 
$m_i$, smoothing length $h_i$, position ${\bf r}_i$, velocity 
${\bf v}_i$, and values for any other local intensive 
thermodynamic variables required to describe the state of the gas, 
for example density $\rho_i$, pressure $P_i$, specific internal 
energy $u_i$, magnetic field ${\bf B}_i$, etc. 

At an arbitrary position ${\bf r}$, the value of any physical 
variable $A$ can be obtained by interpolating over the values 
of $A$ associated with nearby particles, using a smoothing 
kernel $W$, viz.

\begin{equation}
A({\bf r}) \; = \; \sum_i \left\{ \frac{A_i m_i}
{\rho_i h_i^3} \, W\left( \frac{|{\bf r} - {\bf r}_i|}
{h_i} \right) \right\} \; .
\end{equation}

\noindent The kernel is normalized, 

\begin{equation}
\int_{s=0}^{s=\infty} \, W(s) \, 4 \pi s^2 ds \; = \; 1 \; ,
\end{equation}

\noindent and has compact support (specifically $W(s>2) = 0$), so that 
the sum is over a finite number ${\cal N}\nei$ of neighbouring 
particles.

The gradient of $A$ can then be evaluated from 

\begin{equation}
\nabla A({\bf r}) \, = \, \sum_i \left\{ \frac{A_i m_i}
{\rho_i h_i^4} \, W'\left( \frac{|{\bf r} - {\bf r}_i|}
{h_i} \right) \, \frac{{\bf r} - {\bf r}_i}
{|{\bf r} - {\bf r}_i|} \right\} \; ,
\end{equation}

\noindent where $W'(s) \equiv dW/ds$. 

The motions of the particles are driven by interparticle forces 
representing the local pressure gradient, viscosity, gravity 
and magnetic field. Interparticle forces are formulated 
symmetrically so that linear and angular momentum are conserved. 
In our basic implementation we use the M4 kernel introduced by 
Monaghan \& Lattanzio (1985). The particles all have the same 
mass. We omit magnetic fields.

The smoothing length $h_i$ is varied so that the kernel contains 
${\cal N}\nei \sim 50 \; (\pm 5)$ particles. This ensures that the 
spatial resolution becomes finer with increasing density.

The density at the position of particle $i$ is given by

\begin{equation}
\label{SPHDENSITY}
\rho_i \; = \; \sum_{j} \left\{ \frac{m_i}{\bar{h}_{ij}^3} \, 
W\left( \frac{|\Delta {\bf r}_{ij}|}{\bar{h}_{ij}} \right) \right\} \; ,
\end{equation}

\noindent where

\[
\bar{h}_{ij} \; = \; \frac{h_i + h_j}{2} \; , \;\;\;
\Delta {\bf r}_{ij} \; = \; {\bf r}_i - {\bf r}_j \; ,
\]

\noindent The sum in Eqn. (\ref{SPHDENSITY}) is over the $\sim {\cal N}\nei$ 
neighbours within $2 \bar{h}_{ij}$, and it includes particle $i$ itself. 

In many applications, the pressure at particle $i$ is given by a barotropic 
equation of state $P_i = P(\rho_i)$. In this case there is no need to solve 
an energy equation.

The gravitational acceleration of a particle can be obtained by 
a direct sum of the contributions from all the other particles, 

\begin{equation}
\label{GRAV}
{\bf g}_i \; = \; - \, \sum_{j \ne i} \left\{ \frac{G m_j}
{|\Delta {\bf r}_{ij}|^2} \, \frac{\Delta {\bf r}_{ij}}
{|\Delta {\bf r}_{ij}|} \right\} \; .
\end{equation}

\noindent However, there are two problems with this approach. The first 
problem arises because individual particles are very massive. 
As a result, close interactions between particles (small 
$|\Delta {\bf r}_{ij}|$) lead to artificially large accelerations 
which corrupt the simulation. This can be cured by softening the 
mutual gravity of any two particles having small separation. 
The form of softening we use assumes that the kernel describes the 
distribution of a particle's mass, and then invokes Gauss's 
Gravitational Theorem. The effect is to introduce a term 
$W^*(|\Delta {\bf r}_{ij}|/\bar{h}_{ij})$ into the sum of Eqn. 
(\ref{GRAV}) (see Eqn. (\ref{MOTION}), below), where 

\begin{equation}
W^*(s) \; = \; \int_{s'=0}^{s'=s} W(s') \, 4 \pi s'^2 \, ds' \; .
\end{equation}

The second problem arises because the cost of computing the sum 
in Eqn. (\ref{GRAV}) for all the particles increases as $i\ttl^2$ 
(where $i\ttl$ is the total number of particles). Since the 
linear resolution, and the time-step, of an SPH simulation 
are proportional to $i\ttl^{-1/3}$, the cost of increasing the 
resolution soon becomes prohibitive. An effective solution 
to this problem is to arrange the particles in an hierarchical octal 
tree structure, i.e. a nested grid of cells within cells. This 
is called Tree-Code Gravity (TCG; Barnes \& Hut 1986, Hernquist 
1987).

The top cell of the tree is the entire computational domain, 
assumed here to be a cube. The top cell is then partitioned 
into eight equal cubic sub-cells, these in turn are each 
divided into eight equal cubic sub-sub-cells, and so on. 
Partitioning is continued until the lowest cells in the tree 
contain either one particle or no particle. Then the mass $M_I$, 
centre of mass ${\bf R}_I$, linear size $L_I$ and quadrupole 
moment ${\bf Q}_I$ of each cell $I$ are calculated and stored 
(along with some other cell parameters which are useful for 
finding neighbours).

When the gravitational acceleration of particle $i$ is being 
computed, the contribution from a distant cell can be evaluated 
to sufficient accuracy using its ${M_I,{\bf R}_I,{\bf Q}_I}$, 
and there is no need to consider its constituent particles 
individually. For particle $i$ at ${\bf r}_i$, cell $I$ at 
position ${\bf R}_I$ is sufficiently distant to justify this 
approximation if

\begin{equation}
L_I \; < \; \theta\cri \, |{\bf R}_I - {\bf r}_i| \; ,
\end{equation}

\noindent with $\theta\cri = 0.576$ (Salmon, Warren \& Winckelmans 1994).

The equation of motion for particle $i$ is then

\begin{eqnarray} \nonumber
\label{MOTION}
\frac{d{\bf v}_i}{dt} & = & - \, \sum_j \left\{ 
\frac{m_i}{\bar{h}_{ij}^4} \left[ \frac{P_i}{\rho_i^2} + 
\frac{P_j}{\rho_j^2} + \Pi_{ij} \right] \times \right.  \\ \nonumber
 & & \hspace{2.27cm} \left. 
W'\left( \frac{|\Delta {\bf r}_{ij}|}{\bar{h}_{ij}} \right) 
\frac{\Delta {\bf r}_{ij}}{|\Delta {\bf r}_{ij}|} 
\right\} \\
 & & - \, \sum_{j \ne i} \left\{ \frac{m_j}
{|\Delta {\bf r}_{ij}|^2} W^* \left( \frac{|\Delta {\bf r}_{ij}|}
{\bar{h}_{ij}} \right) \frac{\Delta {\bf r}_{ij}}
{|\Delta {\bf r}_{ij}|} \right\} \; . \\ \nonumber
\end{eqnarray}

\noindent Here, the first sum represents short-range forces (hydrostatic 
and viscous forces) and is only over nearby particles. The 
second sum gives the gravitational acceleration, which is 
long-range, 
and is nominally over all particles. In practice, distant 
particles are grouped into clusters using the TCG algorithm, 
as described in the preceding paragraphs.

$\Pi_{ij}$ is the artificial viscosity term advocated by 
Lattanzio et al. (1985), viz. 

\begin{equation}
\Pi_{ij} \; = \; \left\{ \begin{array}{ll}
\frac{- \alpha \mu_{ij} \bar{c}_{ij} + \beta \mu_{ij}^2}
{\bar{\rho}_{ij}} \; , \hspace{0.5cm} & \mu_{ij} < 0 \; ;\\
 & \\
0 \; , & \mu_{ij} > 0 \; ;\\
\end{array} \right.
\end{equation}

\noindent where 

\begin{equation}
\mu_{ij} \; = \; \frac{\left( \Delta {\bf v}_{ij} \mbox{\Large\bf .} 
\Delta {\bf r}_{ij} \right) \bar{h}_{ij}}
{|\Delta {\bf r}_{ij}|^2 + 0.01 \bar{h}_{ij}^2}  \; ,
\end{equation}

\noindent $\Delta {\bf v}_{ij} \equiv {\bf v}_i - {\bf v}_j$, 
$\bar{c}_{ij} = (c_i + c_j) / 2$, and $\bar{\rho}_{ij} = 
(\rho_i + \rho_j) / 2$. We have used $\alpha = 1$ and $\beta = 1$.

The equation of motion is integrated using a second-order 
Runge-Kutta scheme:

\begin{eqnarray} \nonumber
{\bf a}_i^n & = & 
{\bf a}\left( {\bf r}_i^n, {\bf v}_i^n \right) \; ; \\ \nonumber
{\bf r}_i^{n + \hlf} & = & 
{\bf r}_i^n \, + \, {\bf v}_i^n \frac{\Delta t}{2} \; ; \\ \nonumber
{\bf v}_i^{n + \hlf} & = & 
{\bf v}_i^n \, + \, {\bf a}_i^n \frac{\Delta t}{2} \; ; \\ \nonumber
{\bf a}_i^{n + \hlf} & = & 
{\bf a}\left( {\bf r}_i^{n + \hlf}, 
{\bf v}_i^{n + \hlf} \right) \; ; \\ \nonumber
{\bf r}_i^{n + 1} & = & 
{\bf r}_i^n \, + \, {\bf v}_i^{n + \hlf} \Delta t \; ; \\ \nonumber
{\bf v}_i^{n + 1} & = & 
{\bf v}_i^n \, + \, {\bf a}_i^{n + \hlf} \Delta t \; ; \\ \nonumber
\end{eqnarray}

\noindent where ${\bf r}_i^n$, ${\bf v}_i^n$ and ${\bf a}_i^n$ are 
the position, 
velocity and acceleration of particle $i$ at time-step $n$.

The code invokes a binary hierarchy of multiple particle time-steps, 
$\Delta t^*_k = 2^{-k}\Delta t^*\mxm\,$ with $k = 0, 1, 2, ..., k\mxm=30$. 
For each particle $i$ we evaluate a maximum time-step according to

\begin{equation}
\Delta t_i \; = \; 0.3 \; \mbox{MIN} \left\{ 
\frac{1}{|\nabla {\Large\bf .} {\bf v}|_i} , \, 
\frac{h_i}{|{\bf v}|_i} , \, 
\left( \frac{h_i}{|{\bf a}|_i} \right)^{1/2} , \, 
\frac{h_i}{\sigma_i} 
\right\}
\end{equation}

\noindent where 

\begin{equation}
\sigma_i \; = \; c_i \, + \, 1.2 \left( \alpha \, c_i \, + \, \beta \; 
\mbox{MAX} \left\{ \mu_{ij} \right\} \right) \, .
\end{equation}

\noindent The particle is then given the largest time-step $\Delta t^*_k$ 
from the 
binary hierarchy which is smaller than $\Delta t_i$, i.e.

\begin{equation}
k \; = \; \ell og_2 \left\{ \frac{\Delta t^*\mxm}{\Delta t_i} \right\} \, 
+ \, 1 \, , \hspace{0.5cm} \Delta t_i \; \longrightarrow 
\Delta t^*_k \, .
\end{equation}

A more detailed description of the basic code can be found in 
Turner et al. (1995).

\section{Particle splitting}

The resolution of SPH is of order the local smoothing length $h$, 
and with ${\cal N}\nei \sim 50$, $h \sim (m/\rho)^{1/3}$. Therefore 
in order to improve the resolution of a simulation locally we need 
to reduce the masses of the individual particles, and hence to 
increase the overall number of particles. 

In Particle Splitting, this is done by replacing individual 
particles (parent particles) with small groups of particles 
(families of children particles). In the standard form of SPH, 
an individual particle is spherically symmetric, by virtue of 
having an isotropic smoothing kernel. Therefore we should 
distribute the family of children particles which replace it 
so that the sum of their smoothing kernels is approximately 
spherically symmetric.

This is achieved by creating thirteen children, each with mass 

\begin{equation}
m\cld \; = \; \frac{m\pnt}{13} \; .
\end{equation}

\noindent The children are then placed on an hexagonal close-packed array, 
with the first child at the same position as the parent, and the 
other twelve children equidistant from the first; the method used 
to determine the optimum value for the distance $\ell$ between the 
first child and each of its siblings is described in the next section. 
Finally the array of particles is rotated to an arbitrary orientation.

The choice of 13 children is a compromise between two considerations. 
On the one hand, there should not be too large a disparity between 
parent mass and child mass, otherwise there will be greatly increased numerical 
diffusion wherever parents and children are neighbours. On the other hand, 
it is desirable that the collective density distribution of the family of 
children approximate to the spherically symmetric density distribution of 
their parent. With 13 children there is a relatively small variance in 
the collective density of the children on a spherical surface centred on 
the position of the parent.

\section{Fine-tuning the family of children particles}

The children-particles are initially given smoothing lengths 

\begin{equation}
h\cld \; = \; (13)^{-1/3} h\pnt \; .
\end{equation}

\noindent so that they have the same net volume as their parent-particle.

Two experiments have been preformed to determine the optimum value 
for $\ell$ (the distance between the first child and each of its 
twelve siblings). 

In the first experiment we take a microscopic approach. We consider 
a single parent particle and compute the volume-weighted 
difference between its density and the combined densities of its 
children:

\begin{equation}
\int_{\mbox{\small all space}} \left| \sum_{i'} \left\{ \rho_{i'}({\bf r}) \right\} - \rho\pnt({\bf r}) \right| \, d^3{\bf r} \; ,
\end{equation}

\noindent where the summation index $i'$ represents the identifiers of the 
children. This quantity has its minimum 
value when $\ell \simeq 1.9 h\cld$, but it is still quite large, 
$\sim 0.33 \, m\pnt$.

In the second experiment we take a macroscopic approach. 500 
equal-mass particles are distributed randomly inside a cube. 
They are then settled using SPH with uniform sound speed and 
periodic boundary conditions, until the density is very uniform, 
specifically $\sigma_\rho/\bar{\rho} < 0.01$. Then we split all 
the particles simultaneously and re-settle them. For 
$\ell = 1.5 h\cld$, the value of $\sigma_\rho/\bar{\rho}$ is a 
minimum ($\sim 0.10$) immediately after splitting, and the time 
required to settle back to $\sigma_\rho/\bar{\rho} < 0.01$ is also 
a minimum. The minimum of $\sigma_\rho/\bar{\rho}$ immediately 
after splitting is quite shallow (it increases to $\sim 0.12$ for 
$\ell \simeq 1.2 h\cld$, and for $\ell \simeq 2.0 h\cld$), and this 
is largely testimony to the smoothing ability of the M4 kernel 
used with $\sim 50$ neighbours (Monaghan \& Lattanzio 1985).

We believe that this second experiment is the more relevant, since 
it is the macroscopic properties of the fluid which SPH seeks to mimic. 
In other words, it is the continuous density and velocity fields 
(which at most positions are compounded by contributions from $\sim 
{\cal N}\nei$ particles) that SPH must aspire to reproduce, and not 
the density of a single parent-particle.

All subsequent tests are therefore performed with $\ell = 1.5 h\cld$.

\section{Nested Splitting}

One way to implement Particle Splitting is to start a standard 
simulation at $t=0$, then at $t=t\spl (\geq 0)$ to identify 
a part of the overall computational domain, i.e. a sub-domain, where 
improved resolution is required or desired, and to split all 
particles in this sub-domain. In the sequel we refer to the 
initial standard simulation and the particles in it as -- respectively -- the 
coarse simulation and the coarse particles. We refer to the 
simulation after $t\spl$ and the split particles in the sub-domain 
as -- respectively -- the fine simulation and the fine particles.

At time $t=t\spl$ the newly-created fine child-particle $i'$ is 
given a velocity by summing contributions 
from the $\sim 50$ neighbours (index $j$) of -- and including -- 
the coarse parent-particle (index $i$):

\begin{equation}
{\bf v}_{i'} \; = \; \sum_j \left\{ \frac{m_j {\bf v}_j}
{\rho_j \bar{h}_{ij}^3} \, W\left( \frac{|{\bf r}_{i'} - 
 {\bf r}_j|}{\bar{h}_{ij}} \right) \right\} \, .
\end{equation}

\noindent Thereafter ($t > t\spl$), any coarse parent-particle 
which enters the sub-domain is immediately split and the 
resulting fine child-particles are again given 
velocities by summing over the neighbours of the 
coarse parent-particle. The densities and accelerations 
of fine particles are calculated using the standard SPH 
equations (Eqns. (\ref{SPHDENSITY}) \& (\ref{MOTION})).

Near the boundary of the sub-domain, there are coarse particles 
which have fine particles as neighbours, and vice versa. Because 
of this, we have to revise the way in which the smoothing lengths 
of particles are adjusted. The results are greatly improved if, 
instead of requiring the number of neighbours of particle $i$ to 
be $\sim 50$, we require the total mass of the neighbours of 
particle $i$ to be $\sim 50$ times the mass of particle $i$, i.e.

\begin{equation}
\label{NEIB}
\sum_j \left\{ m_j \right\} \; \sim 50 m_i \; .
\end{equation}

\noindent For fine particles well inside the sub-domain, and for coarse 
particles well outside the sub-domain, this makes no difference, 
but it is important for particles near the boundary, including 
coarse particles which are about to be split.

\section{On-The-Fly Splitting}

A second way to implement Particle Splitting is to identify 
particles whose resolution, $\sim (m/\rho)^{1/3}$, is insufficient 
for the problem in hand, and split them on-the-fly. To do this, we 
must recognize what the deterministic physics of the problem is, 
and quantify -- as a local function of state -- the minimum 
resolution required to capture this physics. Then, whenever the 
resolution of a particle approaches the local minimum 
resolution, that particle is split. 

As in Nested Splitting, the velocities of the new 
children-particles are evaluated by summing contributions from 
the neighbours of the parent particle. Successive generations 
of splitting are possible, yielding particles with masses 
$m/13$, $m/13^2$, $m/13^3$, etc. and hence linear resolution improved 
by factors 2.4, 5.5, 13, etc.

As in Nested Splitting, better smoothing is obtained if the 
desired number of neighbours is evaluated according to Eqn. 
(\ref{NEIB}).

In the context of star formation, gravitational collapse and 
fragmentation are crucial physical processes, and therefore the 
code must always be able to resolve the Jeans mass. This is called 
the Jeans Condition (Truelove et al. 1997, Bate \& Burkert 1977). 
Since the Jeans mass is a local function of 
state, it is straightforward to formulate the Jeans Condition, 
and to split particles when they are about to violate it. We 
explore On-The-Fly Particle Splitting triggered by the Jeans 
Condition, in the next two sections.

\section{The Jeans Condition}

Truelove et al. (1997) have shown that simulations of 
protostellar collapse and fragmentation performed using AMR 
only converge when the local Jeans length $\lambda\Jea \sim 
a (G\rho)^{-1/2}$ is resolved (where $a$ is the local sound 
speed). Specifically the linear resolution $\Delta r$ must 
everywhere satisfy 

\begin{equation}
\label{JCAMR}
\Delta r \; \stackrel{<}{\sim} \; \frac{\lambda\Jea}{4} \; .
\end{equation}

\noindent Otherwise artificial fragmentation can occur and/or real 
fragmentation can be suppressed. Eqn. (\ref{JCAMR}) is the 
formulation of the Jeans Condition appropriate for FD codes.

Bate \& Burkert (1997) have shown that there is a similar 
Jeans Condition for SPH simulations: the minimum resolvable 
mass $\sim 2 {\cal N}\nei m$ must everywhere be less than 
the local Jeans mass $\sim G^{-3/2} \rho^{-1/2} a^3$, or

\begin{equation}
\label{JCSPH}
\frac{G^3 \rho_i m_i^2}{a_i^6} \; < \; 
\frac{1}{(2{\cal N}\nei)^2} \; \sim \; 10^{-4} \; .
\end{equation}

\noindent Otherwise artificial fragmentation may occur, and real 
fragmentation will definitely be suppressed.

Whitworth (1998) has shown that provided the smoothing kernel 
is sufficiently centrally peaked, SPH simulations of collapse 
are unlikely to suffer from artificial fragmentation. However, 
it remains the case that Eqn. (\ref{JCSPH}) 
must be satisfied if real fragmentation is to be modelled.

\section{Rotating collapse test}

A standard test of star formation codes is the one proposed 
by Boss \& Bodenheimer (1979; hereafter BB79). In this test 
the initial conditions are a uniform-density isothermal cloud 
having  mass $M_\odot$, radius $0.02\,$pc, density $2 \times 
10^{-18}$ g cm$^{-3}$ and isothermal sound speed $0.17 $ km 
s$^{-1}$ (so the ratio of thermal to gravitational energy is 
$\sim\,$0.26). The cloud then has an $m = 2$ azimuthal density 
perturbation with 10\% amplitude imposed on it, and it is set 
to rotate at uniform angular speed $7.2 \times 10^{-13}$ rad 
s$^{-1}$ (so the ratio of rotational to gravitational energy 
is $\sim\,$0.16). There is no external pressure, and the gas 
remains isothermal during the subsequent evolution, until the 
density rises above a critical value $\rho\cri$, when it switches 
to being adiabatic and therefore heats up.

Klein et al. (1999) have performed this test, with $\rho\cri 
= 5 \times 10^{-14}$ g cm$^{-3}$, using AMR, and shown that the 
cloud should collapse to produce an elongated structure. This 
structure then evolves into a binary system with a bar between 
the components. As long as the gas remains isothermal, the 
binary components and the inter-connecting bar condense into 
filamentary singularities, and the bar does not fragment 
(Truelove et al. 1998). These 
results agree with the theoretical analysis of 
Inutsuka \& Miyama (1992), but contrast with earlier simulations 
of the BB79 test using both FD and SPH codes. In 
particular, the simulations reported by Burkert \& Bodenheimer 
(1993) resulted in the bar fragmenting. The reason for this appears 
to have been that their simulations did not satisfy the Jeans 
Condition.

Bate \& Burkert (1997) have subsequently repeated the BB79 test 
with $\rho\cri = 10^{-14}$ g cm$^{-3}$, using SPH and 80,000 
particles, sufficient to satisfy the Jeans Condition up to 
a density $\sim \rho\cri$. They now find essentially the same 
result as Truelove et al. (1998) and Klein et al. (1999), 
confirming that the earlier results of Burkert \& Bodenheimer (1993) 
were compromised by violating the Jeans condition. Thus 
the BB79 test appears to be a stringent test of whether a code 
is satisfying the Jeans Condition.

We have performed three SPH simulations of the BB79 test. We have 
required the gas to remain isothermal up to a density $\rho\cri = 
5 \times 10^{-12}$ g cm$^{-3}$. This is two orders of 
magnitude higher than the value 
used by Klein et al. (1999), and so in our simulations the 
Jeans mass and length fall to values 10 times smaller, i.e. 
$M\Jea \sim 10^{-4} M_\odot$ and $R\Jea \sim 2$ AU. Thus we 
are subjecting our code to an even sterner test than that advocated 
by Klein et al. (1999). If one accepts the notion of opacity-limited 
fragmentation, then in nature the Jeans mass should not fall below $\sim 
10^{-3} M_\odot$. Therefore -- in this limited sense 
-- we are also subjecting our code to a sterner test than 
that set by nature. 

In the standard simulation there are 600,000 
particles from the outset, sufficient to satisfy the Jeans 
condition without Particle Splitting. In the other two simulations 
there are only 45,000 particles at the outset, and the Jeans condition 
is accommodated by implementing particle splitting, first Nested 
Particle Splitting and then On-The-Fly Particle Splitting. The 
Nested Particle Splitting simulation ends up with 
$\sim$ 320,000 particles, and the On-The-Fly Particle Splitting 
simulation ends up with $\sim$ 140,000 particles.

We follow the suggestion of Truelove et al. (1997, 1998) that numerical results 
should be compared, not at precisely the same elapsed time, but 
instead when the peak density in the computational domain, 
$\rho\pea$, reaches the same 
value. In particular we present results when $\rho\pea \simeq \rho\cri 
= 5 \times 10^{-12}$ g cm$^{-3}$,  
and when $\rho\pea \simeq 2 \times 10^{-9}$ g cm$^{-3}$. The reason 
for this is that the evolution of a simulation appears to 
progress slightly more rapidly if the resolution is higher. The 
same effect was reported by Truelove et al. (1997, 1998) for their 
AMR simulations.

The plots are all grey-scale column-density images through 
the central 0.004 pc $\times$ 0.004 pc (800 AU by 800 AU) 
of the computational domain, viewed down the rotation axis. 
The calibration of the grey-scale is given in the figure 
captions. Times are given in terms of the initial freefall time 
$t\FF \simeq 0.03$ Myr.

\begin{figure*}
\label{STANDARD}
\setlength{\unitlength}{1mm}
\begin{picture}(80,90)
\includegraphics{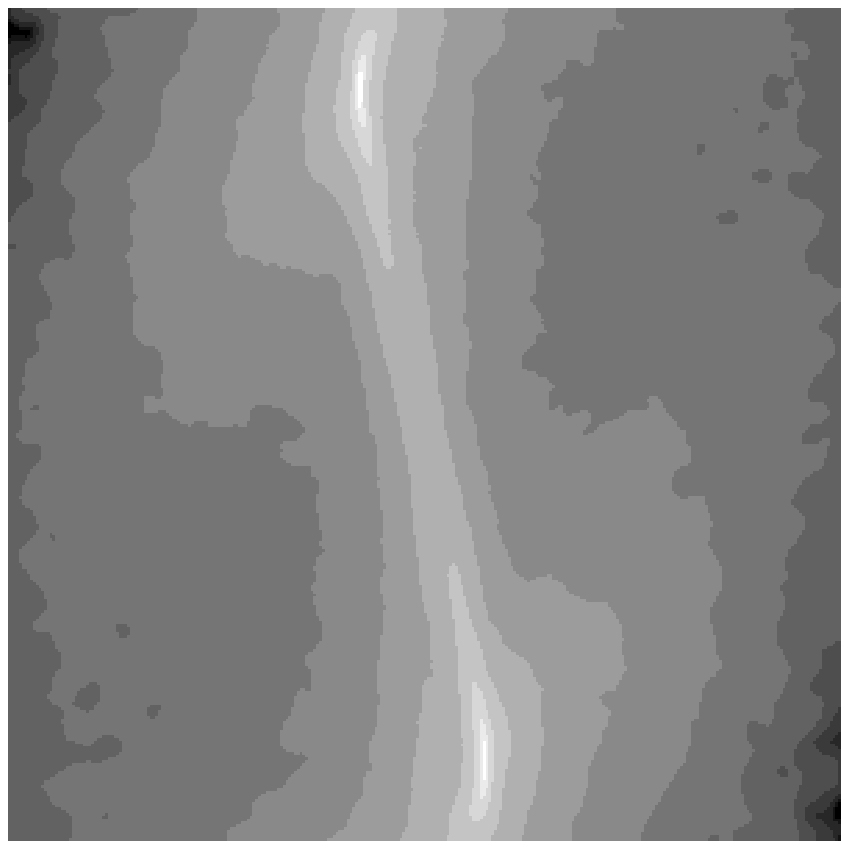}
\end{picture}
\begin{picture}(80,90)
\includegraphics{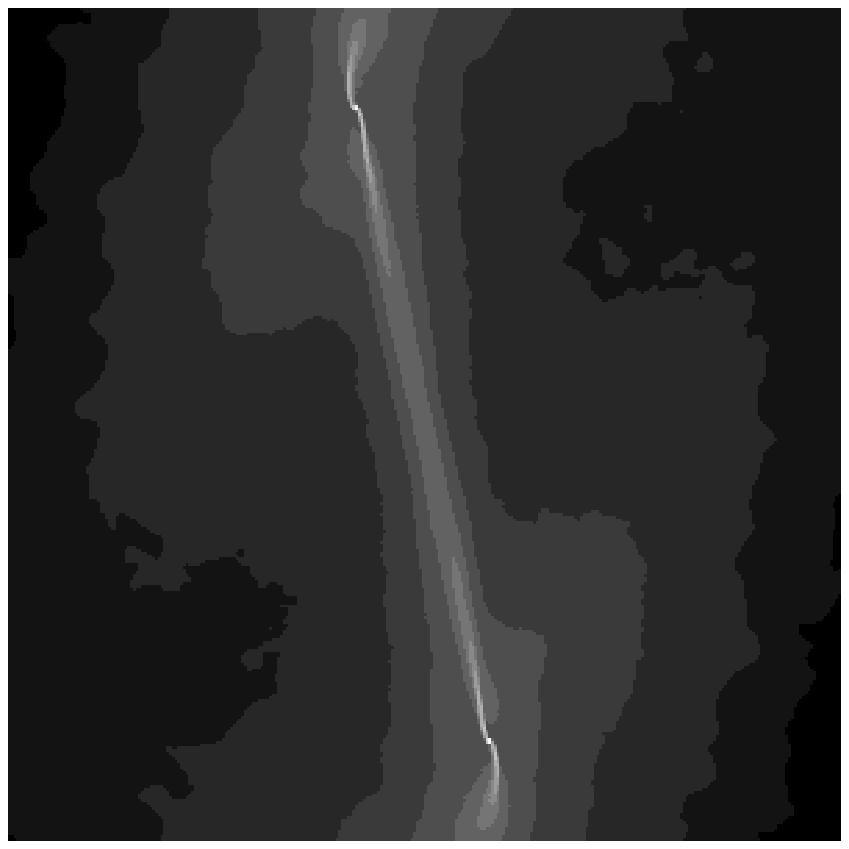}
\end{picture}
\caption{The BB79 test performed using standard SPH with 600,000 
constant-mass particles and anticlockwise rotation: column-density 
images of the central $0.004 \times 0.004$ pc$^2$ of the computational 
domain, looking down the rotation axis. (a) At time $t = 1.237 t\FF$, 
when $\rho\pea \simeq \rho\cri = 5 \times 10^{-12}$ g cm$^{-3}\,$; 
the grey-scale is logarithmic, with sixteen equal intervals from 
$3.60 \times 10^{23} \, {\rm H}_2 \, {\rm cm}^{-2}$ to 
$4.56 \times 10^{25} \, {\rm H}_2 \, {\rm cm}^{-2}$ (or equivalently, 
adopting solar composition, $1.44 \, {\rm g} \, {\rm cm}^{-2}$ to 
$1.82 \times 10^2 \, {\rm g} \, {\rm cm}^{-2}$). (b) At time 
$t = 1.245 t\FF$ when $\rho\pea \simeq 2 \times 10^{-9}$ g cm$^{-3}\,$; 
the grey-scale is logarithmic, with sixteen equal intervals from 
$3.72 \times 10^{23} \, {\rm H}_2 \, {\rm cm}^{-2}$ to $1.55 \times 
10^{27} \, {\rm H}_2 \, {\rm cm}^{-2}$ (or equivalently $1.49 \, 
{\rm g} \, {\rm cm}^{-2}$ to $6.19 \times 10^3 \, {\rm g} \, {\rm cm}^{-2}$).}
\end{figure*}

\subsection{Standard high-resolution SPH simulation \\ (no Particle Splitting)}

In Figure 1 we show the results of a simulation of the BB79 test 
with $\rho\cri = 5 \times 10^{-12}$ g cm$^{-3}$, performed using 
our standard SPH code with 600,000 constant-mass particles. This simulation 
has sufficient particles to satisfy the Jeans condition throughout 
the simulation without Particle Splitting. We note that our SPH 
code has been developed entirely independently of that used by 
Bate \& Burkert (1997) and differs from that code in several 
major regards, in particular the integration scheme and the 
gravity solver.

Figure 1a shows a grey-scale column-density 
image of the centre of the computational domain at the moment 
the density in the binary components $\rho\pea$ passes $\rho\cri$, 
i.e. at $t = 1.237 t\FF$. Both the binary components and the bar 
connecting them approximate to filamentary singularities, and 
there is no sign of bar fragmentation. 

Figure 1b shows the 
same region at the end of the simulation, $t = 1.245 t\FF$. 
By this stage the gas in the densest parts, i.e. the binary 
components, has become adiabatic and heated up. As a result 
the binary components are approximately circular in projection. 
However, the interconnecting bar is still isothermal; it 
continues to approximate to a filamentary singularity and 
shows no sign of fragmenting.

We re-iterate that this simulation extends the isothermal 
evolution to $\rho\cri = 5 \times 10^{-12}$ g cm$^{-3}$, 
which is two orders of magnitude higher than the 
critical density used by Klein et al. (1999). Therefore it 
constitutes a very stern test of the code's validity.

\begin{figure*}
\label{NESTED}
\setlength{\unitlength}{1mm}
\begin{picture}(80,90)
\includegraphics{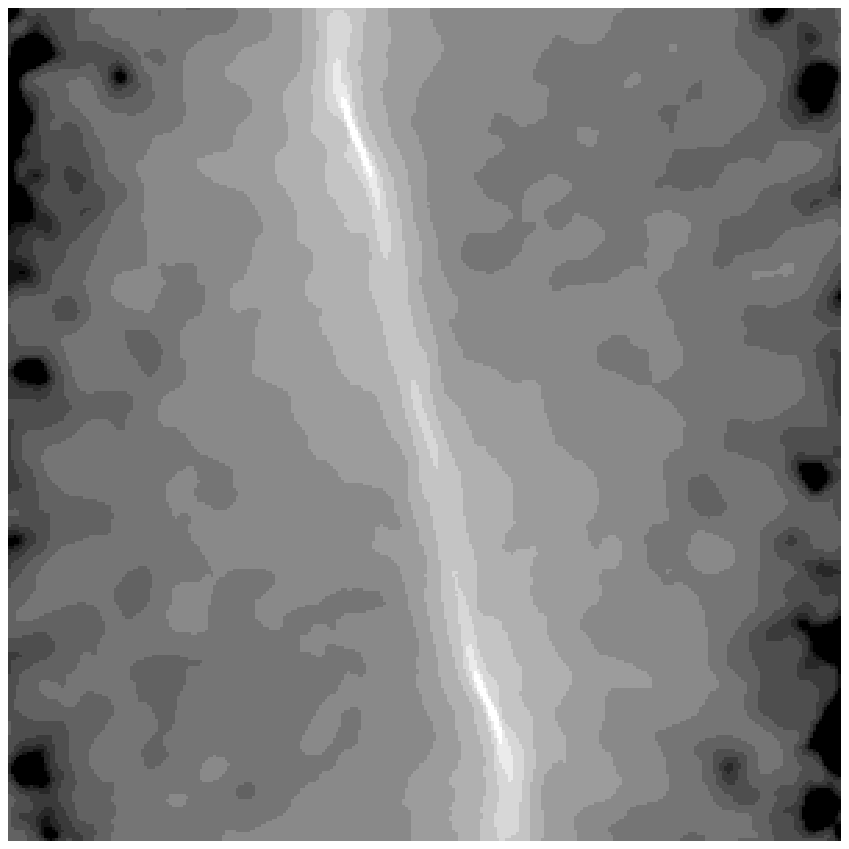}
\end{picture}
\begin{picture}(80,90)
\includegraphics{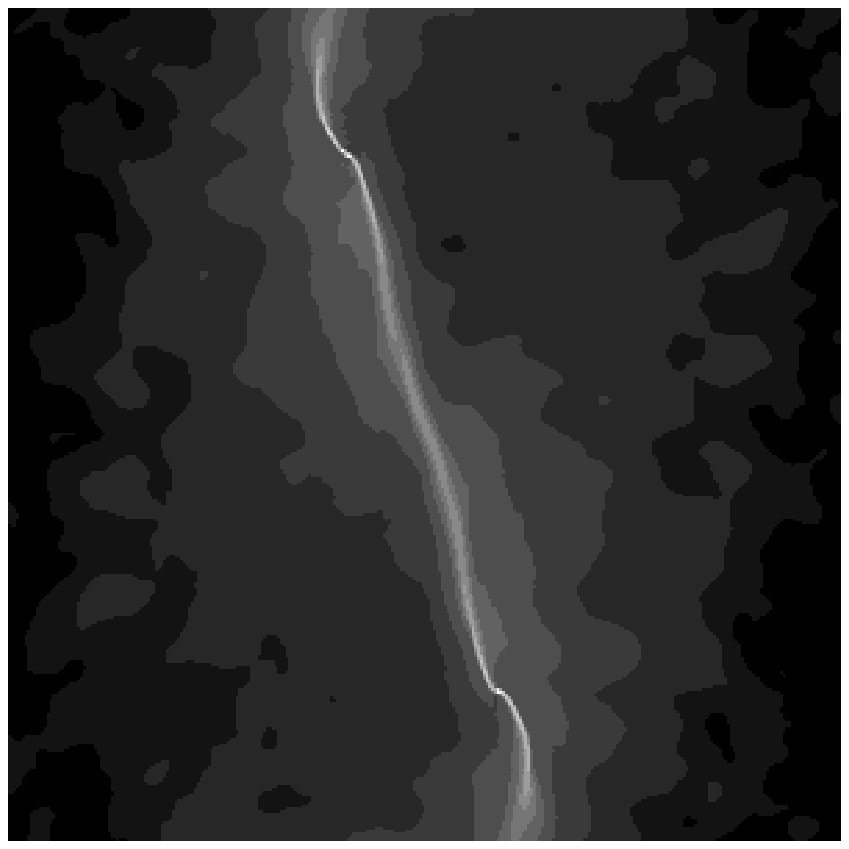}
\end{picture}
\caption{As Fig. 1, but performed using only 45,000 particles 
initially and then Nested Particle Splitting in a central cylindrical 
sub-domain of radius 0.003 pc and height 0.003 pc. (a) At time 
$t = 1.258 t\FF$, when $\rho\pea \simeq \rho\cri = 5 \times 
10^{-12}$ g cm$^{-3}\,$. (b) At time $t = 1.265 t\FF$ when 
$\rho\pea \simeq 2 \times 10^{-9}$ g cm$^{-3}\,$.}
\end{figure*}

\subsection{SPH simulation with Nested Particle Splitting}

Figure 2 shows the result of a simulation of the BB79 test, 
again with $\rho\cri = 5 \times 10^{-12}$ g cm$^{-3}$, 
but now performed with only 45,000 equal-mass particles at the 
outset. Nested Particle Splitting is applied after $t_{\rm split} = 1.244 t\FF$, 
within a cylindrical sub-domain of radius 0.003 pc (600 AU) 
and height 0.003 pc (600 AU). The symmetry axis of the cylinder 
is aligned with the rotation axis of the cloud. Its size is 
chosen so as to contain all regions which become so dense that 
without Particle Splitting they would eventually violate the 
Jeans Condition, i.e. the final binary components, which 
have an orbit of radius 0.002 pc (400 AU).

Figures 2a and 2b 
correspond to times $t = 1.258 t\FF$ (when $\rho\pea \simeq \rho\cri$) 
and $t = 1.265 t\FF$ (the end of the simulation at $\rho\pea \simeq 
2 \times 10^{-9}$ g cm$^{-3}$). By comparing Figs. 1 and 2, we 
see that the simulation with Nested Particle Splitting reproduces the 
crucial features of the evolution found in the standard high-resolution 
simulation described in Section 8.1. 
As before, a binary system forms with a bar between the 
components, and as long as the gas remains isothermal the 
binary components and the bar evolve towards filamentary 
singularities, with no tendency for additional fragments to 
condense out of the bar. There is a small density peak at the 
centre of the bar in the first frame (2a; $\rho\pea \simeq 
\rho\cri$), but this quickly disperses, and therefore we do 
not consider it to be critical.

By the end, approximately half the particles have been split, 
so there are $\sim$ 320,000 particles in total. The simulation 
with Nested Particle Splitting therefore requires less than 
half the memory and approximately 25\% of the CPU used by the 
standard simulation (with 600,000 particles, \S 8.1).

\begin{figure*}
\label{ONTHEFLY}
\setlength{\unitlength}{1mm}
\begin{picture}(80,90)
\includegraphics{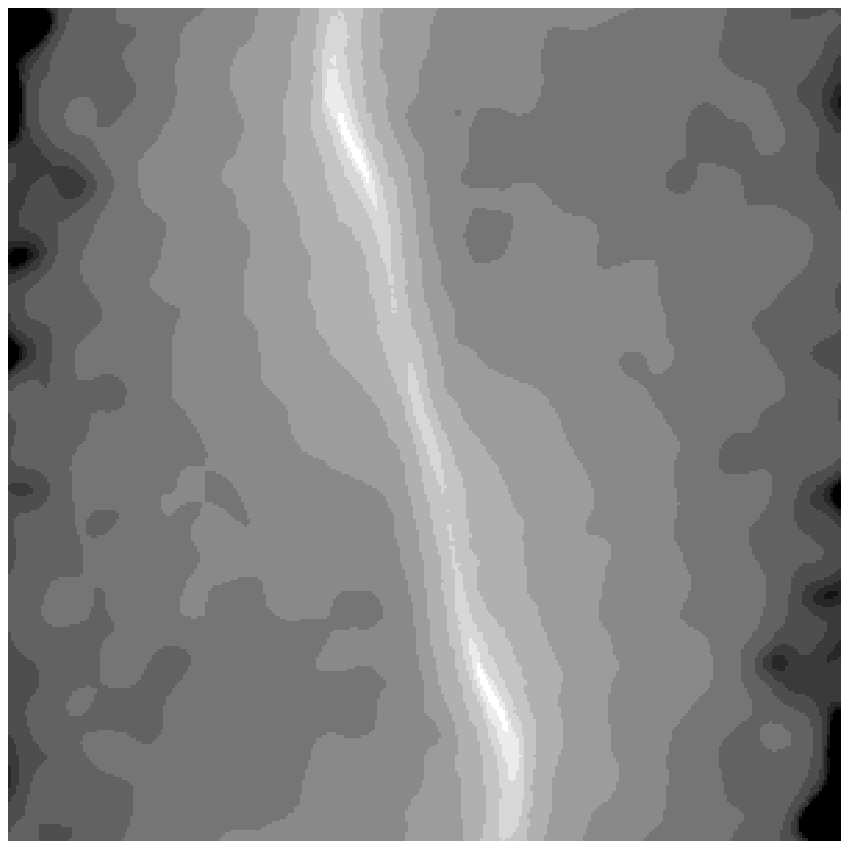}
\end{picture}
\begin{picture}(80,90)
\includegraphics{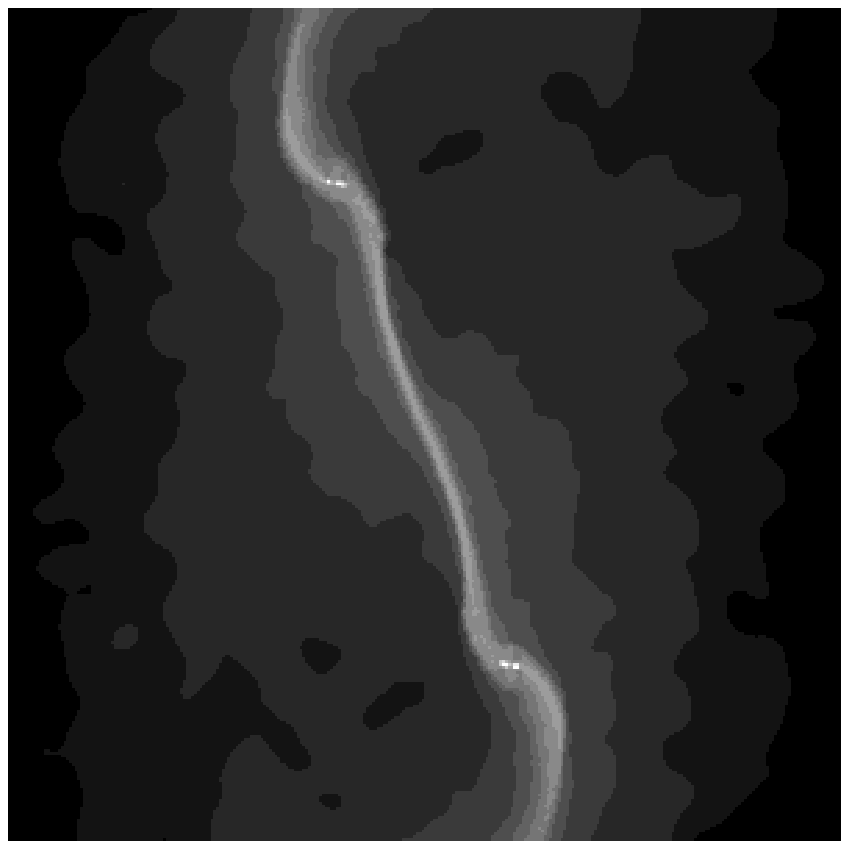}
\end{picture}
\caption{As Fig. 1, but performed using only 45,000 particles 
initially and then On-The-Fly Splitting triggered above $\rho = 
\rho_{\rm split} = 3 \times 10^{-14}$ g cm$^{-3}$. (a) At time $t = 1.259 
t\FF$, when $\rho\pea \simeq \rho\cri = 5 \times 10^{-12}$ g 
cm$^{-3}\,$. (b) At time $t = 1.270 t\FF$ when $\rho\pea \simeq 
3 \times 10^{-9}$ g cm$^{-3}\,$.}
\end{figure*}

\subsection{SPH simulation with On-The-Fly Particle Splitting}

Finally Fig. 3 shows the result of a simulation of the BB79 
test, again with $\rho\cri = 5 \times 10^{-12}$ g cm$^{-3}$, 
and again performed using only 45,000 equal-mass particles 
at the outset, but now with  Particle Splitting applied 
On-The-Fly, in response to imminent violation of the Jeans 
Condition.

Figs. 3a and 3b correspond to times $t = 1.259 t\FF$ (when 
$\rho\pea \simeq \rho\cri$) and $t = 1.270 t\FF$ (the end of the 
simulation at $\rho\pea = 3 \times 10^{-9}$ g cm$^{-3}$; a 
slightly higher final density than in the other two simulations, see
below). Again the simulation reproduces all the critical features 
reported by Truelove et al. (1998) and Klein et al. (1999), 
in particular no tendency for additional fragments to 
condense out of the bar. Once more there is a small density peak 
at the centre of the bar in the first frame (3a; $\rho\pea \simeq 
\rho\cri$), which then quickly disperses.

The only significant difference is a fanning-out of the bar close 
to the binary components, right at the end of the simulation. 
This is a real effect, due to the tidal 
shearing of the bar by the inspiralling binary components. 
It only appears in this simulation because the evolution has 
been followed to somewhat higher density ($3 \times 10^{-9} \, 
{\rm g} \, {\rm cm}^{-3}$ instead of $2 \times 10^{-9} \, {\rm g} \, 
{\rm cm}^{-3}$); hence the binary components have gotten further 
out of alignment with the bar, where they can exert a disruptive 
shear on the ends of the bar.

At the end, less than one fifth of the original particles have 
been split, so there are $\sim$ 140,000 particles in total. 
The simulation with On-The-Fly Splitting therefore requires 
less than one quarter the memory and less than 10\% 
of the CPU used by the standard simulation  (with 600,000 particles, \S 8.1).

\section{Discussion}

It appears that Particle Splitting is a viable option for 
increasing the resolution, locally, during an SPH simulation of 
self-gravitating collapse. In this context, 
the most stringent and apposite test of the algorithm is the BB79 
test, where extra resolution is required to avoid violating the Jeans 
Condition (and hence to avoid artificial fragmentation). Acceptable 
results are obtained for the BB79 test, 
even when the gas is programmed to stay isothermal up to 
$\rho\cri = 5 \times 10^{-12}$ g cm$^{-3}$. Perturbations 
due to the splitting process are transient, and are effectively 
and quickly damped. Moreover, the simulation is accomplished with 
much less memory and CPU 
than a standard simulation, particularly with the On-The-Fly 
implementation. We note also that the savings in memory and CPU 
are likely to be far greater in realistic simulations of star formation, 
where only a fraction of the matter in a cloud collapses 
to form stars, and therefore a far smaller fraction of particles 
needs to be split and to be followed with a small time-step.

Other tests have been performed using the two Particle 
Splitting algorithms. In particular we note the following. (i) 
We have evolved a static, stable isothermal sphere, 
contained by an external pressure $P_{\mbox{\small ext}}$, 
and shown (a) that it 
remains stable, and (b) that its density profile is not 
corrupted. The configuration treated has mass $M_\odot$, 
radius $0.01$ pc, and temperature 7.9 K. It is quite centrally 
condensed, with $\rho_{\mbox{\small central}} \simeq 3 \, 
\rho_{\mbox{\small edge}}$ and $P_{\mbox{\small ext}} \simeq 
3 \times 10^{-12}$ erg cm$^{-3}$. (ii) 
We have also followed the collapse of an isothermal cloud 
which initially has uniform density, and quickly evolves 
towards an inverse-square density profile ($\rho \propto 
r^{-2}$). The cloud has mass $M_\odot$ and temperature 7.9 K. 
Initially it has radius 0.016 pc and uniform density $\sim 
4 \times 10^{-18}$ g cm$^{-3}$. The effect of Particle Splitting 
on the ensuing collapse is negligible.

Particle Splitting might be applied in other situations where 
increased resolution is required locally. The Nested implementation 
requires prior knowledge of where extra resolution will be 
required. The On-The-Fly implementation requires that the condition 
for requiring extra resolution can be formulated as a local 
function of state. However, more development and testing is needed 
to establish the stability of the method in these more general 
situations.

One difference between Particle Splitting and AMR is that 
Particle Splitting -- at least in its present form -- is 
irreversible. If fluid flows into, and then out of, a 
region where extra resolution is required, particles get 
split and stay split. However, there is no a priori reason 
why particles should not be merged 
in regions where they are delivering much higher resolution 
than is required by the local physics, and indeed this was done 
successfully by both Monaghan \& Varnas (1988) and by Meglicki, 
Wickramasinghe and Bicknell (1993), using a grid. We have not yet 
attempted to do this, so we cannot comment further on its 
feasibility.

The concern is that, without merging, if a single low-mass 
particle is surrounded 
by high-mass particles, Eqn. (\ref{NEIB}) dictates that it has 
very few neighbours (typically only 4 if the surrounding particles 
are 13 times more massive). Therefore the low-mass particle might 
have rather noisy density, pressure. etc. because of small-number 
statistics and sudden changes in its neighbour list. However, this 
will be to a large extent offset by two factors. First, by implication 
the fluid is well resolved by the large particles, so they have a 
well relaxed distribution and the neighbourhood of the low-mass 
particle should be quite constant. Second, as regards the 
macroscopic continuum properties of the fluid, within the sphere of 
influence of the low-mass particle (i.e. within its smoothing kernel) 
it makes only a $\sim 2 \%$ contribution to the local thermodynamic 
variables, and so fluctuations in its individual properties are 
very diluted.

Shocks and ionization 
fronts can be treated effectively using ellipsoidal kernels 
(Owen et al. 1998; Francis 2001). Moreover, in the context of star 
formation, irreversibility may not be a serious problem, because 
once gravitational collapse and fragmentation are underway the 
local Jeans mass is unlikely to increase much, and so the resolution 
required by the Jeans Condition is unlikely to decrease. We are 
currently developing a version of our code which treats magnetic fields 
and ambipolar diffusion (Hosking \& Whitworth, in preparation), but 
until this has been fully tested without particle splitting it would 
not be sensible to combine the two.

\section {Conclusions}

We have described how Particle Splitting can be implemented in SPH, 
and we have shown that it offers a stable and economic way to 
increase the local resolution of an SPH simulation of self-gravitating 
collapse, and hence 
avoid violating the Jeans Condition. This makes Particle Splitting 
a useful refinement for SPH simulations of gravitational collapse 
and fragmentation.

\section*{Acknowledgements}

The authors thank Henri Boffin and Neil Francis for many interesting 
discussions on this project.

\end{document}